# Toward Air-Stable Multilayer Phosphorene Thin-Films and Transistors

Joon-Seok Kim[*], Yingnan Liu[*], Weinan Zhu[*], Seohee Kim, Di Wu, Li Tao, Ananth Dodabalapur, Keji Lai, and Deji Akinwande

\* These authors contributed equally to this work.

AFFILIATIONS

1. Microelectronics Research Center, Department of Electrical and Computer Engineering, The University of Texas at Austin, Austin, TX, 78758, USA.
Joon-Seok Kim, Weinan Zhu, Seohee Kim, Li Tao, Ananth Dodabalapur, Deji Akinwande

2. Department of Physics, The University of Texas at Austin, Austin, TX, 78712, USA.
Yingnan Liu, Di Wu, Keji Lai

CORRESPONDING AUTHORS
Keji Lai, kejilai@physics.utexas.edu; Deji Akinwande, deji@ece.utexas.edu

**ABSTRACT**

Few-layer black phosphorus (BP), also known as phosphorene, is poised to be the most attractive graphene analogue owing to its high mobility approaching that of graphene, and its thickness-tunable band gap that can be as large as that of molybdenum disulfide. In essence, phosphorene represents the much sought after high-mobility, large direct band gap two-dimensional layered crystal that is ideal for optoelectronics and flexible devices. However, its instability in air is of paramount concern for practical applications. Here, we demonstrate air-stable BP devices with dielectric and hydrophobic encapsulation. Microscopy, spectroscopy, and transport techniques were employed to elucidate the aging mechanism, which can initiate from the BP surface for bare samples, or edges for samples with thin dielectric coating highlighting the ineffectiveness of conventional scaled dielectrics. Our pioneering months-long studies indicate that a double layer of $Al_2O_3$ and hydrophobic fluoropolymer affords BP devices and transistors with indefinite air-stability for the first time, overcoming a critical material challenge for applied research and development.

**INTRODUCTION**

Bridgman discovered van der Waals bonded black phosphorus crystals in 1914 during high-pressure experiments[1]. While much of its bulk semiconducting properties have been investigated for one hundred years[2], it is only recently that few-layer black phosphorus or phosphorene, an elemental two-dimensional (2D) layered crystal, has risen to scientific limelight[3-5]. As a physical analogue to the graphite system, black phosphorus has a distinct puckered or buckled structure[2]. Its fully satisfied valence shell results in a direct band gap[2,3], which has been found to scale inversely with thickness, around 1.5-2 eV for a monolayer and 0.3 eV in the bulk limit[4,6]. Notably, BP's high mobility (about 1000 $cm^2/V{\cdot}s$)[3] and sizeable band gap place it at the electronic intersection of graphene, a zero-gap high mobility 2D crystal[7,8], and semiconducting transitional metal dichalcogenides such as $MoS_2$, a large-gap low mobility 2D crystal[9,10]. In addition, the thickness-tunable direct band gap of BP is ideal for optoelectronics spanning the infrared to visible regimes, with a favorable potential for hyperspectral photodetection[11,12].

A fundamental material challenge for BP films is the lack of air-stability[13-17], a matter of paramount importance for devices that typically operate in ambient conditions. Similar to hygroscopic red phosphorus[1,18], it is found that unprotected black phosphorus can absorb moisture upon exposure to air, resulting in compositional and physical changes of the material with consequent degradation of its electronic properties. This poses a severe problem for prospective device applications in semiconductor technology and flexible electronics[7-9,19-23]. To date, active research efforts continue to take place to obtain a complete understanding of the degradation process and a pathway towards air-stable BP devices[13,14,16,17]. A number of recent studies have implemented dielectric capping layers to provide stability for material and device research[16,24-28].

In this light, we have systematically investigated the stability of unprotected and capped black phosphorus flakes mechanically exfoliated onto insulating substrates. In our study, BP experienced physical degradation in a matter of hours in air, as determined by atomic force microscopy (AFM) and optical imaging, and complete device failure in a few days. Microwave impedance microscopy with nanoscale spatial resolution revealed that samples with thin capping layers, though more air-stable, began to degrade from the edges inward within a few days, a unique mechanism different from previously reported surface degradation[13,17]. Intriguingly, the degradation was primarily electronic with only minor changes in flake thickness and volume, in sharp contrast to unprotected samples. Our results indicate for the first time that visual or topographical techniques are generally inadequate for monitoring the air-degradation in capped BP. Moreover, statistics from BP field-effect transistors (FETs) have shown that double layer capping with dielectric and fluoropolymer films afford robust months-long air-stability, suggesting indefinite long-term stability beyond the timescale of hours and weeks reported in previous studies[13,14,16,17,28]. The effectiveness of the double layer capping is attributed to the fluoropolymer's hydrophobicity, which prevents moisture from being adsorbed and diffusing to the black phosphorus interface. The simplicity of the capping methods represents a facile route for achieving air-stable black phosphorus or phosphorene devices that can enable basic studies and potential applications.

**RESULTS**

Multilayer BP samples were exfoliated onto 25 nm thick $Al_2O_3$ on Si substrates. Flakes of thicknesses between 5 and 25 nm were selected by optical and atomic force microscopy. For back-gated FETs, device patterning was done by electron-beam lithography and electrical contacts were formed by electron-beam evaporation of 2/70 nm Ti/Au (**Fig. 1a**). The optical image of a typical BP device with contact electrodes is shown in **Fig. 1b**. Raman spectra of typical exfoliated BP flakes showed the three characteristic vibrational modes[4] (**Fig. 1c**).

**Exposed black phosphorus**. Uncapped BP samples under the normal lighted in-door ambient conditions (temperature 24–27 ºC, relative humidity 40–45 %) experienced material degradation in a matter of hours, which can be seen in the optical and AFM images of a representative flake with smooth terraces of different thicknesses in **Fig. 1d**. This observation of short term degradation of BP upon exposure to air has been also reported in recent studies[13-17]. Additional AFM data collected within 24 hrs indicated moisture accumulation on the surface, which resulted in substantial local volume expansion and a cleated surface with sharp spikes. A video clip showing AFM data acquired over one week is available as a **Supplementary Media File**. Corresponding optical images showed increasing optical transparency during the process, indicative of continuous thinning of the flake. As the surrounding $Al_2O_3$ surface area showed no discernible moisture adsorption, we conclude that the degradation is localized to the black phosphorus flake. Electrical measurements on a similar flake showed substantial degradation over two days and complete device failure in a week (**Fig. 1e**), with the flake vanishing from the substrate surface (see **Supplementary Note 1**).

The degradation and swelling of uncapped BP is considered to be a consequence of moisture absorption, an attribute of hygroscopic materials[29]. The absorbed moisture has two adverse effects on BP films: i) physical changes such as volume expansion and cleated surfaces (**Fig. 1d**), and ii) chemical changes towards a liquid phase, which eventually vanishes from the surface (**Supplementary Note 1**). While the precise composition and stoichiometry of the liquid phase is unknown, energy dispersive x-ray spectroscopy (EDAX) indicates that the droplet-like cleats or swells are oxygen rich (**Supplementary Note 2**), which may suggest phosphorus oxides in line with recent studies[16,17,30], or phosphorus oxyacids as suggested in the early work of Bridgman[1]. Moreover, Raman spectrum of a degraded flake (**Supplementary Note 3**) displayed peaks characteristic of P-O stretching modes[31-35], lending further experimental support to the suggested oxides or oxyacids. The vast diversity of phosphorus compounds[36], owing to its wide range of oxidation states (+1 to +5) and coordination numbers, and the possible influence of native impurities or defects in the BP crystal warrant rigorous chemistry research beyond the scope of this work to precisely characterize the air-degraded products.

**Thin-cap black phosphorus**. In order to improve the air-stability of BP films, a thin (~2-3 nm) layer of aluminum was deposited on the exfoliated flakes and oxidized into $Al_2O_3$ [25,28,37], to serve as a barrier to moisture adsorption, as depicted in **Fig. 2a**. This range of small dielectric thickness is technologically relevant for scaled 2D FETs[38]. In addition, a thin passivation layer here also entails that the dielectric is thinner than the multilayer BP flakes. To obtain nanoscale electrical information on the thin-cap BP over an extended period of time, an AFM-based microwave impedance microscope (MIM) was employed (**Fig. 2a**) to simultaneously map out the surface topography and electrical properties with a spatial resolution of 10~100 nm[39,40]. In the MIM experiment, a 1 GHz signal is guided to the tip apex of a customized cantilever probe and the reflected signal is demodulated by the MIM electronics to form local permittivity and conductivity maps of the material. The excitation microwave power is kept low (~10μW) to avoid non-linear or destructive effects on the sample. Details of the MIM technique are included as **Supplementary Note 4**. Furthermore, a unique feature of MIM is the near-field capacitive coupling, which enables direct electrical imaging of buried structures with a spatial resolution independent of the free-space wavelength. Compared with other surface-sensitive electrical probes such as conductive AFM or scanning tunneling microscope, the MIM's sub-surface imaging capability is particularly useful for the investigation of capped BP samples in our study, and is generally suitable for the electrical monitoring of spatial or temporal variations in thin films[40,41].

In order to understand the effect of surface passivation, we measured a BP sample with thin dielectric capping every day over a period of one week by MIM/AFM (raw data shown in **Supplementary Note 5**). As discussed in **Supplementary Note 5**, the local sheet resistance, which is displayed in **Fig. 2b**, can be estimated from the MIM data using numerical analysis[42]. The simultaneously acquired AFM images are shown in **Fig. 2c**. As seen from the MIM data, the sample right after the dielectric coating was electrically homogenous except for the folded segment at the bottom. The measured sheet resistance $R_{sh} \sim 10^6 \, \Omega/sq$ is also consistent with our transport data (see below) and reported results at zero gate bias[13]. Strikingly, after one day, the local resistance around the sample perimeter decreased sharply by two orders of magnitude, likely due to doping from the environment, with virtually no change in the surface topography. In other words, unlike uncapped films, the initial degradation of BP with thin dielectric capping is mostly electronic rather than structural. The edge initiated process suggests that a capping layer thinner than the flake may offer poor sidewall coverage (**Fig. 2a**), hence relatively ineffective against moisture adsorption at the edges. Moreover, degradation initiated from the sample edge dominates the overall degradation process, further suggesting that conformal capping layer with fully covered sidewalls is essential in providing sufficient encapsulation. For longer exposure time in the ambient, the highly conductive regions propagated toward the center of the flake. At the same time, the local sheet resistance at the edges gradually increased. After one week, the entire flake became electrically uniform again ($R_{sh} \sim 10^5 \, \Omega/sq$), which is still an order of magnitude more conductive than the initial state. At the same time, the AFM data showed appreciable increase of flake thickness at the edges (see **Fig. 3d**).

Raman intensity maps collected after the AFM/MIM experiment (at day 9) displayed the three characteristic modes of BP (**Fig. 2d** and **Supplementary Note 6**), which corroborates the topographical data regarding the physical preservation of the film. Interestingly, a relatively small and broad bump likely composed of several new Raman modes can be seen in the spectrum at higher wavenumbers (800–900 cm$^{-1}$), which is confined to the BP flake. Compared with published literature[20,21,35], this new Raman bump is consistent with the vibrations of mixed phosphorus oxide compounds. Similar new Raman modes were previously observed in doped multi-wall carbon nanotubes as a result of strong dopant-carbon interactions[43]. Based on the AFM/MIM and Raman data, we attribute the degradation of thin-cap BP to partial oxidation initiated from the edges, which drastically impacts the electrical properties while largely preserving the overall physical integrity. With further understanding on their origins, the 800–900 cm$^{-1}$ Raman shifts may become of practical utility in evaluating the purity of BP or phosphorene films similar to the use of the Raman D-mode in characterizing disorder in graphene[44,45].

**Thick-cap black phosphorus**. For improved air-stability, a capping layer thicker than the BP flake was deposited on the as-exfoliated samples. As depicted in **Fig. 3a,** such a thick dielectric coating (~25 nm) is expected to serve as an effective barrier against moisture adsorption at both the surface and edges. Similar thickness effect of an ALD layer was previously used to achieve conformal coatings on pristine carbon nanotubes[46] and BP[24-28,35]. The simultaneous topographic and electrical imaging data of a thick capped BP sample are shown in **Fig. 3b** and **3c** (raw data and numerical simulation in **Supplementary Note 6**). In contrast to the observed electrical inhomogeneity on the thin-cap sample, neither the local sheet resistance nor the surface topography showed discernible changes during the same week-long MIM/AFM monitoring. For a comparison between the two capped samples, selected AFM line profiles at the beginning and the end of the monitoring period are plotted in **Fig. 3d**. The absence of changes in the MIM/AFM data signifies the preservation of the physical, electrical, and chemical integrity of BP under thick capping over the week-long duration.

**Aging of black phosphorus devices**. BP field-effect transistors with different capping schemes were fabricated to elucidate the aging effect at the device level. Electrical characteristics of representative back-gated phosphorene FETs with thin and thick ALD capping are presented in **Fig. 4a** and **4b**, respectively. Detailed sample preparation procedures are described in the Methods section. From **Fig. 4a**, it is evident that the key device parameters, such as the ON current and ON/OFF current ratio, fluctuated substantially for the thin-cap FET over the course of two weeks. The thick-cap FET, on the other hand, showed better stability over two weeks with the overall ON/OFF switching attributes preserved (**Fig. 4b**). This increased stability is in agreement with recent studies on passivation of BP with ALD deposited $Al_2O_3$[28,35]. However, even thick-capped devices with sidewalls fully covered by the ALD metal oxide dielectric showed some extent of degradation, with the main temporal changes comprising of a negative shift of $I_d$-$V_g$ transfer characteristics and an increasingly severe hysteresis over time, which may be attributed to the slow diffusion of adsorbates through the $Al_2O_3$ capping layer over an extended time-scale. This indicates that regardless of the thickness, $Al_2O_3$ alone might be inadequate in providing long-term electrically stable BP devices operating in ambient conditions.

To further improve the air-stability of BP transistors, we have developed a double-layer capping scheme by spin-casting a transparent hydrophobic fluoropolymer film (Teflon-AF)[47-49] on top of the first layer of 25 nm $Al_2O_3$. The hydrophobicity is crucial for preventing moisture adsorption on the device, resulting in much improved air-stability over the same two-week duration (**Fig. 4c**). Further insights on the air-stability of phosphorene FETs with thin, thick and double-layer capping schemes were obtained by analyzing the change in electrical performance metrics as shown in **Fig. 4d-f**. Here, we have collected data on the device performance metrics including ON current, hysteresis in gate voltage, and $I_{on}/I_{off}$ ratio for forty-six BP transistors over a period of three months. We note that the BP samples used in this study are of good electronic quality with initial mobilities around 200 $cm^2/V \cdot s$ and comparable to each other at the beginning of the aging experiment (**Supplementary Note 8**). For thin-cap FETs, severe fluctuations over time and large device-to-device variations are observed, consistent with the strong electrical inhomogeneity observed in the MIM data. $I_{on}/I_{off}$ ratio also dramatically decreased by orders of magnitude within a month. The situation is improved in thick-cap samples, which showed relatively moderate variations especially in hysteresis and $I_{on}/I_{off}$ ratio. However, ON current level experienced large fluctuation and overall decrease in the thick capped devices. In contrast, devices with double-layer capping feature robust performance metrics with minimal variations over time across devices, which we attribute to the combined benefits of physical and chemical protections against moisture and oxygen molecular species.

**DISCUSSION**

The attractive semiconducting properties of reported BP devices with potential mobilities approaching graphene and a direct sizeable band gap similar to monolayer semiconducting transitional metal dichalcogenides have generated substantial interest[3-5,12,13]. However, the poor air-stability of few-layer BP or phosphorene is a major roadblock that precludes the application of bare materials in normal ambient conditions. In order to take advantage of its attractive properties, a carefully designed capping layer is needed to isolate BP from the environment. While an ultra-thin dielectric coating is desirable for maximum gate control or applications in flexible electronics[19,38,50], the present study shows that, without conformal sidewall coverage and moisture resistance from an effective hydrophobic surface, electronic and chemical degradation still occurs and propagates inwards from any exposed edges. This suggests that the aging process of BP may be more complex than originally considered[17,16], with multiple reaction dynamics and pathways contributing to the degradation.

The double-layer coating in this study, which consists of an ALD film followed by the hydrophobic Teflon-AF fluoropolymer, provides a facile route to achieve good air-stability in BP devices. Since direct deposition of fluoropolymers on BP may have adverse effects due to impurity scattering from the solvent residue, the $Al_2O_3$ or other ALD dielectric films are beneficial for ensuring a high quality dielectric-phosphorus interface. We note that since spin-casted Teflon-AF is a transparent thin film[49,51], its utility as a capping layer is applicable to optoelectronic devices without hindering light-matter interactions.

In conclusion, we have systematically investigated the aging process of multilayer BP by microscopy, spectroscopy, and transport measurements. While bare BP experiences severe physical and electrical degradation within a day, carefully engineered ALD dielectric and hydrophobic polymer capping layers afford much improved air-stability over several weeks without compromising the material properties. The non-invasive MIM, with a unique capability of sub-surface conductivity imaging, was employed to elucidate the aging mechanism of ALD-capped BP. For samples with a thin coating layer, electrical degradation signaled by order-of-magnitude change in local resistivity started from the edges and propagated inward, even though the surface topography is largely preserved. Improved air-stability was demonstrated by capping the flake with a thick $Al_2O_3$ layer and the most air-stable BP devices were achieved with an Al$_2$O$_3$ coating followed by a hydrophobic fluoropolymer film. Importantly, this work not only provides deep insights on the aging process of BP but also points to a viable route towards robust BP or phosphorene devices, a practical requirement for prospective applications in nanoelectronics, optoelectronics and flexible electronics.

**METHODS**

**Sample Preparation:**

BP samples were mechanically exfoliated from bulk black phosphorus (Smart Elements) using conventional scotch-tape method, and transferred onto Al$_2$O$_3$/Si substrates in ambient cleanroom conditions. The substrate was prepared by depositing 25 nm-thick ALD Al$_2$O$_3$ on heavily doped p$^{++}$ Si wafer, where the thickness of the dielectric was selected to provide optical contrast for estimation of BP thickness. The Al$_2$O$_3$ layer was also used as the back-gate dielectric in electrical measurements. Samples used in MIM experiments were capped after mechanical exfoliation by either the evaporation of ~2-3 nm Al followed by oxidation in 120 °C for 20 minutes(thin capping), or ALD deposition of 25 nm Al$_2$O$_3$ at 250 °C (thick capping). Back-gated transistor devices were prepared by patterning resist layer using Carl Zeiss FE NEON 40 SEM system, followed by metal evaporation of 2/70 nm Ti/Au. All devices were designed to have channel length of 1 μm. PMMA EL6 and PMMA A3 were spun at 4000 revolutions per minute (rpm) and baked at 180°C for 2 minutes each in sequence to provide initial capping of BP flakes and to form the resist layer. Thin and thick capping of the transistor devices were formed by ALD deposition of either 5 nm or 25 nm Al$_2$O$_3$. Double-layer capping was formed by spin-casting DuPont Teflon-AF at 4000 rpm on regular 25 nm ALD Al$_2$O$_3$ capping layer and cured at 250 °C for 30 minutes. The time of exposure to ambient air before applying capping layers was controlled to be less than 1 hours for all samples.

**AFM and Optical Characterization:**

Few-layer black phosphorus flakes were exfoliated onto $Al_2O_3$/Si substrates as described above and identified using optical microscope. Flakes with thickness from 5 to 25 nm were targeted in this study. AFM images were collected from a Veeco Dimesion 3100 Atomic Force Microscope under the tapping mode. Non-coated n-doped Si TESP probe was used to achieve sub-nanometer scale AFM scanning resolution. AFM images were captured in ambient conditions. In between AFM scans, the monitored sample was kept in ambient environment with in-door lighting. Raman spectra for typical black phosphorus flakes with thickness of 5, 10 and 15 nm were measured using Renishaw Raman system using a green 532 nm laser with fixed polarization angle. Under 100× magnification and with good focus, the laser beam is 1.2 to 1.5 μm in diameter. Raman shift of ~1 $cm^{-1}$ spectrum resolution was achieved using 2400 l/mm gratings.

**MIM Aging Experiments:**

Microwave impedance microscopy (MIM) based on contact-mode AFM was employed to measure topography and local electrical properties simultaneously. Regular AFM images were collected simultaneously with laser feedback from an XE-70 Park AFM. The 1 GHz microwave excitation with $V_{1GHz} \sim 20$ mV was guided through the electrically shielded cantilever (commercially available from PrimeNano Inc.) to the metallic tip with a radius of ~100 nm. The two output channels of MIM correspond to the real and imaginary parts of the local sample admittance, from which the conductivity and/or permittivity of sample can be deduced.

Numerical simulation of the MIM signals was performed by the finite-element analysis (FEA) software COMSOL4.3. The temperature and humidity inside the AFM cabinet were around 25 °C and 46.7 %, respectively. Samples were stored under similar environment during the week-long experiment. More information about the MIM tip can be found from Ref. [40].

**Transistor Device Measurements:**

All electrical measurements were done under ambient conditions in a Cascade summit 11000 AP probing station/Agilent 4156C system using 19 μm-radius tungsten tips from Cascade. The sweep range of back-gate bias was either ±3 or ±5 V, and drain voltage was set to -100mV with source electrode grounded. The drain bias sweep range was from 0 to -3 V, with varying back-gate voltages. Samples were stored at room temperature at atmospheric pressure, 42~46 % humidity and ambient light between each measurements. All measured data was batch-analyzed with Matlab to extract consistent statistic results over diverse devices. On-current values were evaluated at the same overdrive voltage for all devices in order to eliminate current level fluctuations due to changes in the threshold voltages. Hysteresis was calculated from the difference between forward/backward gate sweeps at the maximum $g_m$ values.

**ACKNOWLEDGEMENTS**

This work is supported in part by the Army Research Office under contract W911NF-13-1-0364, the Office of Naval Research (ONR) under contract N00014-1110190, and the Southwest Academy of Nanoelectronics (SWAN) sponsored by the Semiconductor Research Corporation (SRC). The microwave imaging work was supported by Welch Foundation Grant F-1814. D.A acknowledges the TI/Jack Kilby Faculty Fellowship.

**CONTRIBUTIONS**

J.-S.K., K.L and D.A. conceived the original ideas of this study. J.-S.K., W.Z. and L.T. performed material analysis and measurements. Y.L., D.W. and K.L. developed the MIM setup, and performed the measurements and analysis. J.-S.K. and W.Z. prepared and performed measurements of transport samples. J.-S.K. conducted statistical analysis of electrical devices. S.K. and A.D. helped with preparation and analysis of devices with fluoropolymers. D.A. and K.L. led the writing of the paper, and all the authors participated in the discussion of results. The whole project was supervised by A.D., K.L. and D.A.

**FINANCIAL INTERESTS**
The authors declare no competing financial interests

# REFERENCES

1        Bridgman, P. W. Two New Modifications of Phosphorus. *Journal of the American Chemical Society* **36**, 1344-1363, doi:10.1021/ja02184a002 (1914).

2        Morita, A. Semiconducting black phosphorus. *Appl Phys A* **39**, 227-242, doi:10.1007/bf00617267 (1986).

3        Li, L. *et al.* Black phosphorus field-effect transistors. *Nat Nano* **9**, 372-377 (2014).

4        Liu, H. *et al.* Phosphorene: An Unexplored 2D Semiconductor with a High Hole Mobility. *ACS Nano* **8**, 4033-4041 (2014).

5        Liu, H., Du, Y., Deng, Y. & Ye, P. D. Semiconducting black phosphorus: synthesis, transport properties and electronic applications. *Chemical Society reviews*, doi:10.1039/c4cs00257a (2014).

6        Tran, V., Soklaski, R., Liang, Y. & Yang, L. Layer-controlled band gap and anisotropic excitons in few-layer black phosphorus. *Physical Review B* **89**, 235319 (2014).

7        Schwierz, F. Graphene Transistors: Status, Prospects, and Problems. *Proceedings of the IEEE* **101**, 1567-1584 (2013).

8        Wong, H.-S. P. & Akinwande, D. *Carbon Nanotube and Graphene Device Physics*.  (Cambridge Univ Press, 2011).

9        Wang, Q. H., Kalantar-Zadeh, K., Kis, A., Coleman, J. N. & Strano, M. S. Electronics and optoelectronics of two-dimensional transition metal dichalcogenides. *Nat Nano* **7**, 699-712 (2012).

10       Chhowalla, M. *et al.* The chemistry of two-dimensional layered transition metal dichalcogenide nanosheets. *Nat Chem* **5**, 263-275 (2013).

11       Engel, M., Steiner, M. & Avouris, P. A black phosphorus photo-detector for multispectral, high-resolution imaging. *arXiv preprint arXiv:1407.2534* (2014).

12       Buscema, M. *et al.* Fast and Broadband Photoresponse of Few-Layer Black Phosphorus Field-Effect Transistors. *Nano Letters* **14**, 3347-3352, doi:10.1021/nl5008085 (2014).

13       Castellanos-Gomez, A. *et al.* Isolation and characterization of few-layer black phosphorus. *2D Materials* **1**, 025001 (2014).

14       Island, J. O., Steele, G. A., van der Zant, H. S. J. & Castellanos-Gomez, A. Environmental stability of few-layer black phosphorus. *arXiv preprint arXiv: ...* (2014).

15       Koenig, S. P., Doganov, R. A., Schmidt, H., Castro Neto, A. H. & Özyilmaz, B. Electric field effect in ultrathin black phosphorus. *Applied Physics Letters* **104**, 103106, doi:10.1063/1.4868132 (2014).

16       Wood, J. D. *et al.* Effective Passivation of Exfoliated Black Phosphorus Transistors against Ambient Degradation. *Nano letters*, doi:10.1021/nl5032293 (2014).

17       Favron, A. *et al.* Exfoliating black phosphorus down to the monolayer: photo-induced oxidation and electronic confinement effects. *arXiv preprint arXiv:1408.0345* (2014).

18       Silverstein, M. S., Nordblom, G. F., Dittrich, C. W. & Jakabcin, J. J. Stable Red Phosphorus. *Industrial & Engineering Chemistry* **40**, 301-303, doi:10.1021/ie50458a024 (1948).

19       Lee, J. *et al.* 25GHz Embedded-Gate Graphene Transistors with High-K Dielectrics on Extremely Flexible Plastic Sheets. *ACS Nano* **7**, 7744–7750, doi:10.1021/nn403487y (2013).

20       Chang, H.-Y. *et al.* High-Performance, Highly Bendable MoS2 Transistors with High-K Dielectrics for Flexible Low-Power Systems. *ACS Nano* **7**, 5446-5452 (2013).

21       Ramon, M. *et al.* 3GHz Graphene Frequency Doubler on Quartz Operating Beyond the Transit Frequency. *IEEE Transactions on Nanotechnology* **11**, 877 - 883 doi:10.1109/tnano.2012.2203826 (2012).

22       Rahimi, S. *et al.* Toward 300 mm Wafer-Scalable High-Performance Polycrystalline Chemical Vapor Deposited Graphene Transistors. *ACS Nano*, doi:10.1021/nn5038493 (2014).

23       Akinwande, D., Petrone, N. & Hone, J. Two-dimensional flexible nanoelectronics. *Nature Communications*, doi:10.1038/ncomms6678

10.1038/ncomms6678|www.nature.com/naturecommunications (2014).

24      Das, S., Demarteau, M. & Roelofs, A. Ambipolar Phosphorene Field Effect Transistor. *ACS Nano* **8**, 11730-11738, doi:10.1021/nn505868h (2014).

25      Liu, H., Neal, A. T., Si, M., Du, Y. & Ye, P. D. The Effect of Dielectric Capping on Few-Layer Phosphorene Transistors: Tuning the Schottky Barrier Heights. *Electron Device Letters, IEEE* **35**, 795-797, doi:10.1109/LED.2014.2323951 (2014).

26      Liu, H., Neal, A. T. & Ye, P. D. Ambipolar phosphorene field-effect transistors with dielectric capping. *Device Research Conference (DRC), ...* **4133**, 201-202 (2014).

27      Na, J., Lee, Y. T., Lim, J. A., Hwang, D. K. & Kim, G. T. Few-Layer Black Phosphorus Field-Effect Transistors with Reduced Current Fluctuation. *ACS nano* **8**, 11753-11762 (2014).

28      Luo, X., Rahbarihagh, Y., Hwang, J. & Liu, H. Temporal and Thermal Stability of Al2O3-passivated Phosphorene MOSFETs. *Electron Device Letters* **35**, 1314-1316 (2014).

29      Wong, E. H., Rajoo, R., Koh, S. W. & Lim, T. B. The Mechanics and Impact of Hygroscopic Swelling of Polymeric Materials in Electronic Packaging. *Journal of Electronic Packaging* **124**, 122-126, doi:10.1115/1.1461367 (2002).

30      Ziletti, A., Carvalho, A., Campbell, D. K., Coker, D. F. & Neto, A. H. C. Oxygen defects in phosphorene. *arXiv:1407.5880* (2014).

31      Rudolph, W. W. Raman-and infrared-spectroscopic investigations of dilute aqueous phosphoric acid solutions. *Dalton Transactions* **39**, 9642-9653 (2010).

32      Venkateswaran, C. in *Proceedings of the Indian Academy of Sciences, Section A.*   25-30 (Indian Academy of Sciences, 1935).

33      Carbonnière, P. & Pouchan, C. Vibrational spectra for P4O6 and P4O10 systems: Theoretical study from DFT quartic potential and mixed perturbation-variation method. *Chemical Physics Letters* **462**, 169-172, doi:http://dx.doi.org/10.1016/j.cplett.2008.07.056 (2008).

34      Hanwick, T. J. & Hoffmann, P. O. Raman Spectra of Several Compounds Containing Phosphorus. *The Journal of Chemical Physics* **19**, 708-711, doi:doi:http://dx.doi.org/10.1063/1.1748337 (1951).

35      Wood, J. D. *et al.* Effective Passivation of Exfoliated Black Phosphorus Transistors Against Ambient Degradation. *Nano letters* (2014).

36      Fee, D. C., Gard, D. R. & Yang, C.-H. in *Kirk-Othmer Encyclopedia of Chemical Technology* (John Wiley & Sons, Inc., 2000).

37      Kim, S. *et al.* Realization of a high mobility dual-gated graphene field-effect transistor with Al[sub 2]O[sub 3] dielectric. *Applied Physics Letters* **94**, 062107-062103, doi:10.1063/1.3077021 (2009).

38      Lam, K., Dong, Z. & Guo, J. Performance Limits Projection of Black Phosphorous Field-Effect Transistors. *Electron Device Letters, IEEE* **35**, 963-965, doi:10.1109/led.2014.2333368 (2014).

39      Lai, K., Kundhikanjana, W., Kelly, M. & Shen, Z.-X. Nanoscale microwave microscopy using shielded cantilever probes. *Appl Nanosci* **1**, 13-18, doi:10.1007/s13204-011-0002-7 (2011).

40      Yang, Y. *et al.* Batch-fabricated cantilever probes with electrical shielding for nanoscale dielectric and conductivity imaging. *Journal of Micromechanics and Microengineering* **22**, 115040 (2012).

41      Liu, Y. *et al.* Mesoscale Imperfections in MoS2 Atomic Layers Grown by a Vapor Transport Technique. *Nano Letters* **14**, 4682-4686, doi:10.1021/nl501782e (2014).

42      Lai, K., Kundhikanjana, W., Kelly, M. & Shen, Z. Modeling and characterization of a cantilever-based near-field scanning microwave impedance microscope. *Review of scientific instruments* **79**, 063703-063703-063706 (2008).

43      Duclaux, L. Review of the doping of carbon nanotubes (multiwalled and single-walled). *Carbon* **40**, 1751-1764, doi:http://dx.doi.org/10.1016/S0008-6223(02)00043-X (2002).

44      Ferrari, A. C. *et al.* Raman Spectrum of Graphene and Graphene Layers. *Physical Review Letters* **97**, 187401 (2006).

45      Tao, L. *et al.* Synthesis of High Quality Monolayer Graphene at Reduced Temperature on Hydrogen-Enriched Evaporated Copper (111) Films. *ACS Nano* **6**, 2319-2325 doi:http://dx.doi.org/10.1021/nn205068n (2012).

46      Javey, A. *et al.* Carbon nanotube field-effect transistors with integrated ohmic contacts and high-kappa gate dielectrics. *Nano Letters* **4**, 447-450 (2004).

47      Ha, T. J., Lee, J., Akinwande, D. & Dodabalapur, A. The Restorative Effect of Fluoropolymer Coating on Electrical Characteristics of Graphene Field-Effect Transistors. *IEEE Electron Device Letters* **34**, 559-561, doi:10.1109/led.2013.2246537 (2013).

48      Ha, T.-J. *et al.* Transformation of the Electrical Characteristics of Graphene Field-Effect Transistors with Fluoropolymer. *ACS Applied Materials & Interfaces* **5**, 16-20, doi:10.1021/am3025323 (2013).

49      Yang, M. K., French, R. H. & Tokarsky, E. W. Optical properties of Teflon® AF amorphous fluoropolymers. *MOEMS* **7**, 033010-033010-033019, doi:10.1117/1.2965541 (2008).

50      Han, S.-J. *et al.* High-Frequency Graphene Voltage Amplifier. *Nano Letters* **11**, 3690-3693, doi:10.1021/nl2016637 (2011).

51      Lowry, J. H., Mendlowitz, J. S. & Subramanian, N. S. Optical characteristics of Teflon AF fluoroplastic materials. *OPTICE* **31**, 1982-1985, doi:10.1117/12.59910 (1992).

## FIGURES

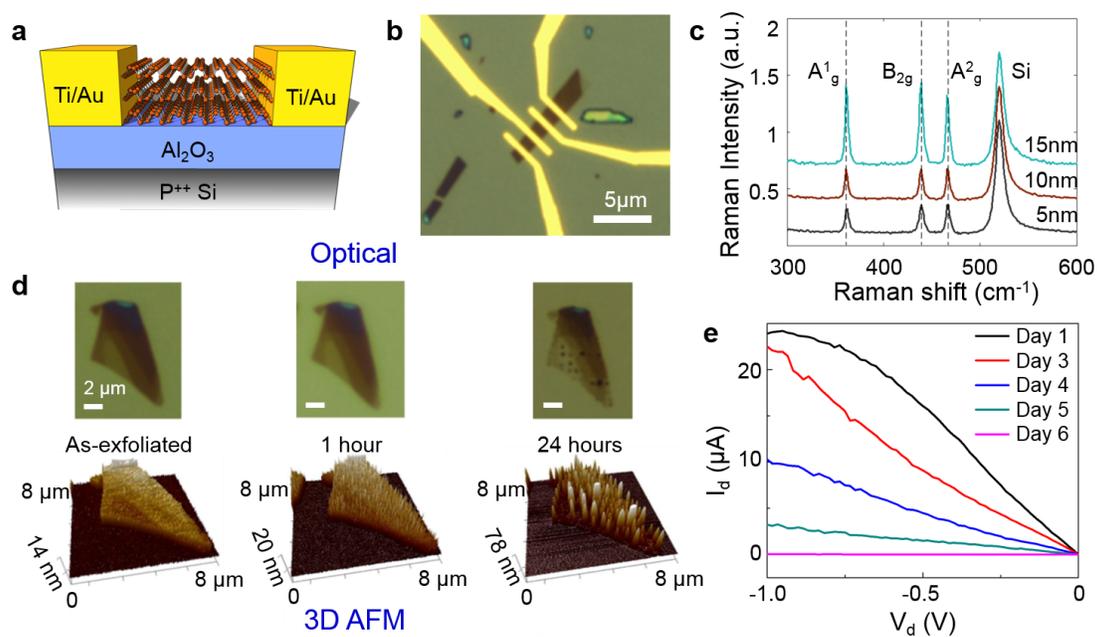

**Figure 1. Characterization of bare black phosphorus.** (**a**) Schematic of back-gated BP FET (not to scale). (**b**) Optical picture of a BP FET device. 25 nm $Al_2O_3$ and 2/70 nm Ti/Au were used as the gate dielectric and electrical contacts, respectively. (**c**) Typical Raman spectra of BP flakes with thicknesses of 5, 10 and 15 nm. (**d**) Optical microscope images (top row) and AFM images (bottom row) illustrating the physical degradation of an exfoliated sample in ambient. (**e**) $I_d$-$V_d$ characteristic of an uncapped BP device, showing severe degradation of electrical conduction over time.

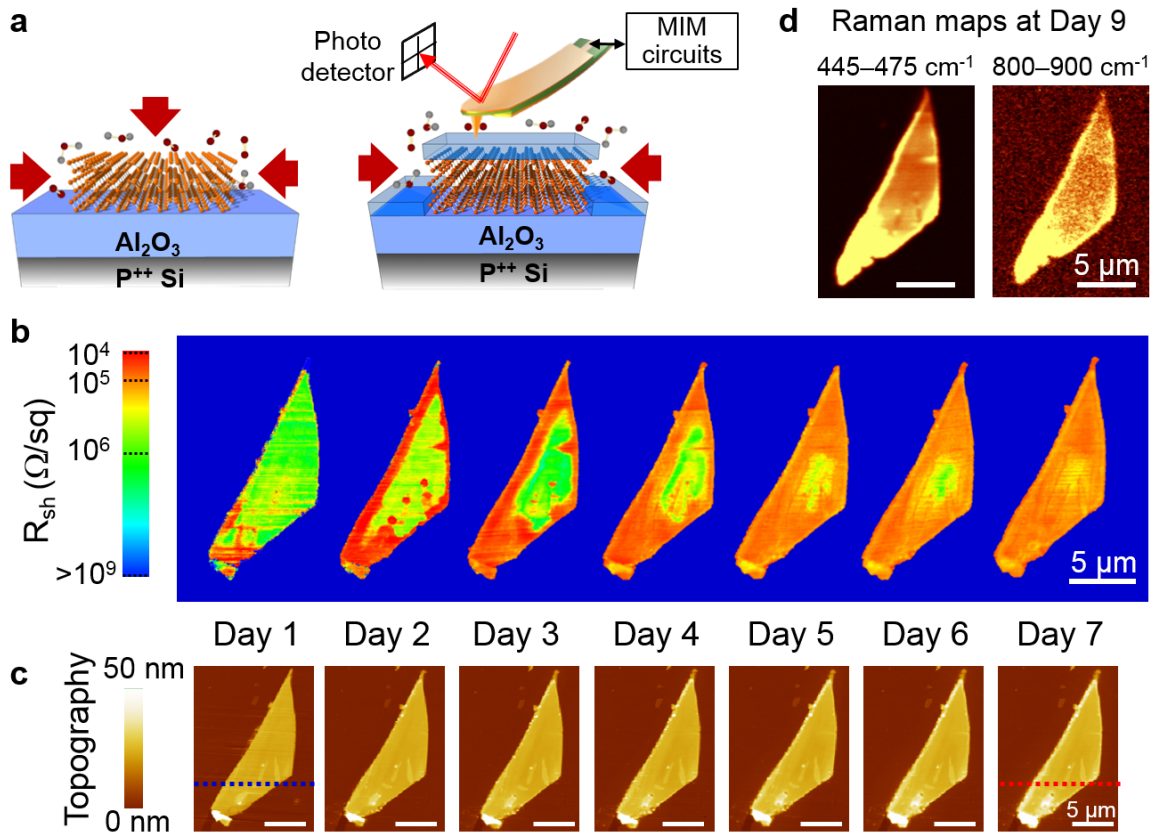

**Figure 2. Spatial and temporal evolution of black phosphorus with thin dielectric capping. (a)** Schematics of BP samples on $Al_2O_3$/Si substrates without (left) and with (right) a thin $Al_2O_3$ capping layer. Red arrows indicate the possible pathways of oxygen and moisture to react with the sample. The MIM setup is also sketched on top of the thin-cap sample. **(b)** Local sheet resistance maps derived from the MIM-Real data of a 24 nm-thick flake capped by $\sim 3$ nm $Al_2O_3$ layer. The images were acquired daily over one week. Significant conductivity changes initiated from the edges are obvious in the images. The high-conductivity regions in the sample interior (day 2 and 3) are likely due to the local thickness variation. **(c)** AFM topography of the same flake simultaneously acquired during the same time duration. The line cuts (blue and red dashed lines) in Day 1 and Day 7 are analyzed in **Fig. 3d**. **(d)** Raman intensity maps integrated over 445–475 $cm^{-1}$ ($A_g^2$ mode of phosphorene) and 800–900 $cm^{-1}$ (consistent with P-O stretching modes) measured at Day 9. All scale bars are 5 μm.

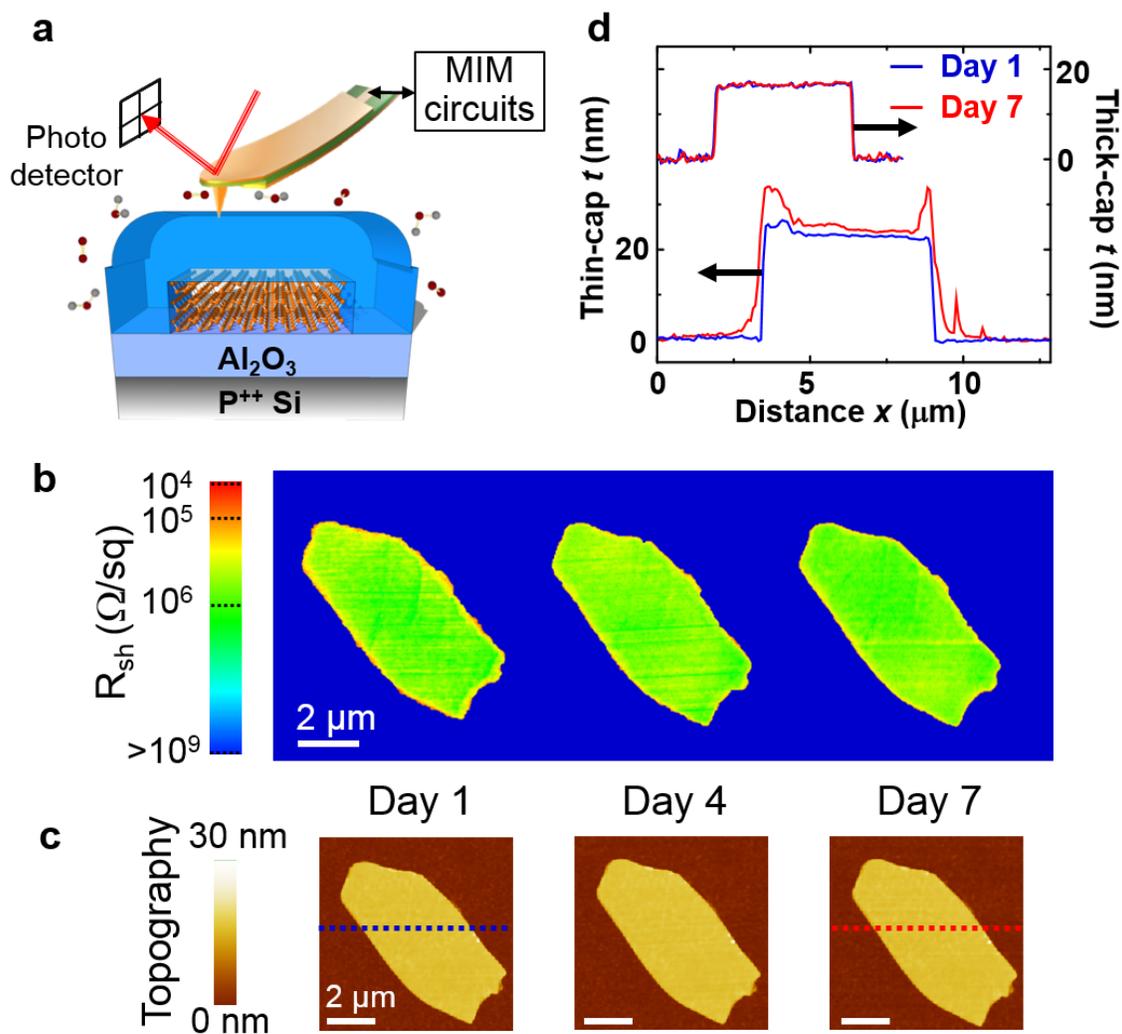

**Figure 3. Spatial and temporal evolution of black phosphorus with thick dielectric capping.** (**a**) Schematic of BP sample on $Al_2O_3$/Si substrate with a thick $Al_2O_3$ capping layer against oxygen and moisture in the ambient. The MIM setup is also illustrated. (**b**) Sheet resistance maps approximated by the MIM-Real data of a 16 nm-thick flake capped by 25 nm $Al_2O_3$. The images were acquired daily over one week. (**c**) AFM topography of the same flake taken during the same time duration. Neither MIM nor AFM showed discernible changes throughout this period. (**d**) AFM line profile of thin-cap (bottom) and thick-cap (top) samples. The blue and red lines are associated with Day 1 and Day 7, respectively. Substantial increase in thickness is seen at the edges of the thin-cap sample, while little change is observed for the sample with thick capping. All scale bars are 2 μm.

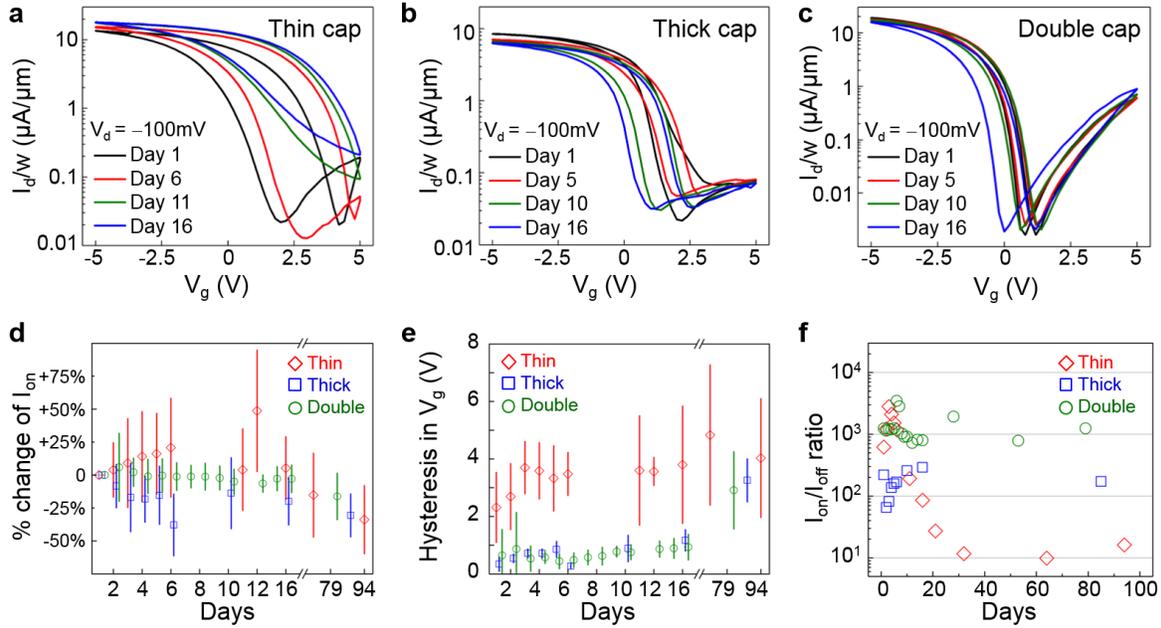

**Figure 4. Aging of black phosphorus FETs.** (**a**-**c**) $I_d$-$V_g$ plot of BP devices with various capping schemes. The drain voltage $V_d = -100$mV for all measurements. (**a**) FETs with thin capping (5 nm ALD-deposited $Al_2O_3$) experienced significant degradation, such as reduced ON/OFF ratio and the increase of gate hysteresis. (**b**) Thick-cap devices (25 nm ALD $Al_2O_3$) showed less aging effect. (**c**) Devices with double-layer capping (25 nm ALD $Al_2O_3$ followed by spin-casting of DuPont Teflon-AF fluoropolymer) showed the least aging effect within the same duration. (**d**-**e**) Statistical analysis of BP FET devices with different capping methods. The mean values of thin-, thick-, and double-cap devices are presented as red diamonds, blue squares, and green circles, respectively. Vertical bars indicate the range of one standard deviation. (**d**) The change of ON current ($I_{on}$) with respect to the values at Day 1. (**e**) Hysteresis in $V_g$ between forward and reverse sweeps. For the two performance metrics shown here, the thin-cap FETs exhibited the largest fluctuations and the best stability is obtained in devices with double-layer capping. (**f**) $I_{on}/I_{off}$ ratio of three selected devices. Thin capped device showed sharp degradation, resulting in $I_{on}/I_{off}$ ratio of ~10x after a month. In contrast, $I_{on}/I_{off}$ ratio of thick capped device was preserved to ~70% compared to Day 1. Double capped device showed the best preservation of $I_{on}/I_{off}$ ratio, with negligible change after 79 days. Initial mobility values of the three selected devices are ~200 cm$^2$/V-s as shown in **Supplementary Note 8**.